\documentstyle[prd,aps,floats,twocolumn]{revtex}

\def\H{{\cal H}}
\def\be{\begin{equation}}
\def\ee{\end{equation}}
\def\bea{\begin{eqnarray}}
\def\eea{\end{eqnarray}}

\newcommand\lsim{\mathrel{\rlap{\lower4pt\hbox{\hskip1pt$\sim$}}
    \raise1pt\hbox{$<$}}}
\newcommand\gsim{\mathrel{\rlap{\lower4pt\hbox{\hskip1pt$\sim$}}
    \raise1pt\hbox{$>$}}}

\newcommand\bfx{{\bf x}}

\begin{document}
\preprint{} 
\draft

%
%
\input epsf
\renewcommand{\topfraction}{0.99}
\renewcommand{\bottomfraction}{0.99}
\twocolumn[\hsize\textwidth\columnwidth\hsize\csname@twocolumnfalse\endcsname

\title{Evolution of Second-Order Cosmological Perturbations
and Non-Gaussianity}
\author{N. Bartolo$^1$, S. Matarrese$^2$ and A. Riotto$^2$}
\address{(1)Astronomy Centre, University of Sussex
Falmer, Brighton, BN1 9QJ, U.K.}
\address{(2) Department of Physics and INFN
Sezione di Padova, via Marzolo 8, I-35131 Padova, Italy}

\date{\today}
\maketitle
\begin{abstract}
\noindent
We present a second-order gauge-invariant formalism to study the evolution 
of curvature perturbations in a Friedmann-Robertson-Walker universe filled 
by multiple interacting fluids.
We apply such a general formalism to describe the evolution of the 
second-order curvature perturbations in the standard one-single
field inflation, in the curvaton and in the inhomogeneous reheating 
scenarios for the generation of the cosmological perturbations.
Moreover, we provide the exact expression for the second-order temperature 
anisotropies on large scales, including second-order gravitational effects 
and  extend the well-known formula for the Sachs-Wolfe effect at linear order. 
Our findings clarify what is the exact non-linearity parameter 
$f_{\rm NL}$ entering in the determination of higher-order statistics such 
as the bispectrum of Cosmic Microwave Background temperature anisotropies. 
Finally, we compute the level of non-Gaussianity in each scenario for the 
creation of cosmological perturbations.

\end{abstract}

\pacs{PACS numbers: 98.80.cq; DFPD-A-03-36}

\vskip2pc]


\section{Introduction}
Inflation \cite{guth81,lrreview} has  become the dominant 
paradigm for understanding the 
initial conditions for structure formation and for Cosmic
Microwave Background (CMB) anisotropy. In the
inflationary picture, primordial density and gravity-wave fluctuations are
created from quantum fluctuations ``redshifted'' out of the horizon during an
early period of superluminal expansion of the universe, where they
are ``frozen'' \cite{muk81,hawking82,starobinsky82,guth82,bardeen83}. 
Perturbations at the surface of last scattering are observable as temperature 
anisotropy in the CMB, which was first detected by the COsmic Background 
Explorer (COBE) satellite \cite{smoot92,bennett96,gorski96}.
The last and most impressive confirmation of the inflationary paradigm has 
been recently provided by the data 
of the Wilkinson Microwave Anistropy Probe ({\it WMAP}) mission which has 
marked the beginning of the precision era of the CMB measurements in space
\cite{wmap1}.
The {\it WMAP} collaboration has  
produced a full-sky map of the angular variations 
of the CMB, with unprecedented accuracy.
{\it WMAP} data confirm the inflationary mechanism as responsible for the
generation of curvature (adiabatic) super-horizon fluctuations. 

Despite the simplicity of the inflationary paradigm, the mechanism
by which  cosmological adiabatic perturbations are generated  is not
established. In the {\it standard scenario} associated
to one-single field models of inflation, the observed density 
perturbations are due to fluctuations of the inflaton field itself. 
When inflation ends, the inflaton oscillates about the minimum of its
potential and decays, thereby reheating the universe. As a result of the 
fluctuations
each region of the universe goes through the same history but at slightly
different times. The 
final temperature anisotropies are caused by the fact that
inflation lasts different amounts of time in different regions of the universe
leading to adiabatic perturbations. Under this hypothesis, 
the {\it WMAP} dataset already allows
to extract the parameters relevant 
for distinguishing among single-field inflation models \cite{ex}.

An alternative to the standard scenario is represented by the {\it curvaton 
mechanism}
\cite{curvaton1,LW,curvaton3,LUW} where the final curvature perturbations
are produced from an initial isocurvature perturbation associated to the
quantum fluctuations of a light scalar field (other than the inflaton), 
the curvaton, whose energy density is negligible during inflation. The 
curvaton isocurvature perturbations are transformed into adiabatic
ones when the curvaton decays into radiation much after the end 
of inflation. 
Recently, another mechanism for the generation of cosmological
perturbations has been proposed \cite{gamma1,gamma2,gamma3}
dubbed the {\it inhomogeneous reheating scenario} (sometimes
called the modulated reheating scenario)\footnote{The idea that the
total curvature perturbation may be affected on large scales
by entropy perturbations when there exists a scalar field affecting
particle masses or couplings constants controlling the reheating
process has first  been suggested in \cite{Qreh}.}.  
It acts during the reheating
stage after inflation if super-horizon spatial
fluctuations in the decay rate of the inflaton field
are induced during inflation, causing  adiabatic perturbations
in  the final reheating temperature
in different regions of the universe.

Contrary to the standard picture, the curvaton and the 
inhomogeneous reheating scenario
mechanism exploit the fact that 
the total curvature perturbation (on uniform density hypersurfaces)
$\zeta$ can change on arbitrarily large scales due to a non-adiabatic
pressure perturbation    which may be 
present  in a multi-fluid system \cite{Mollerach,MFB,GBW,WMLL,wmu}.
While the entropy
perturbations evolve independently of the curvature perturbation on
large scales,  the evolution of the large-scale curvature is
sourced by entropy perturbations. 

Fortunately, the standard and the curvaton  scenarios 
have different observational signatures. 
The curvaton and the 
inhomogeneous reheating scenario
allow to generate the observed level of density perturbations with a 
much lower scale of inflation and thus generically predict 
a smaller level of gravitational waves. 

More interestingly,  the various scenarios for the generation
of the cosmological perturbations predict  different 
levels of non-Gaussianity which 
is usually parametrized by
a dimensionless non-linear parameter $f_{\rm NL}$. 
The non-Gaussianity can be large enough to be 
detectable either by  present  CMB experiments, 
the current {\it WMAP} \cite{k} bound on
non-Gaussianity being  $\vert f_{\rm NL} \vert \lsim 10^2$, or 
by future planned satellites,such as  {\it Planck},
which  have enough resolution to detect 
non-Gaussianity of CMB anisotropy data 
with high precision \cite{ks}.

Since a positive future detection of non-linearity
in the CMB anisotropy pattern might allow
to discriminate among the mechanisms
by which  cosmological adiabatic perturbations are generated, it is
clear that the precise determination of the non-Gaussianity 
predicted by the various  mechanisms is of primary interest.

In this paper we achieve different goals:

\begin{itemize}

\item We present a gauge-invariant formalism 
at second-order 
to study the evolution of curvature perturbations 
in a Friedmann-Robertson-Walker universe filled by 
multiple interacting fluids. We apply this general formalism
to 
describe the evolution of the second-order gauge
invariant curvature perturbations from inflation down to
the radiation/matter phase   in the various
scenarios for the generation of the cosmological
perturbations. In particular, we study the evolution of the
total curvature perturbation during the non-adiabatic phase
when either the inflaton field (for the standard and the
inhomogeneous reheating scenario)
or the curvaton field
oscillate and decay into radiation. 
We show that the total curvature perturbation
is exactly conserved in the standard scenario during the 
reheating phase after inflation and is sourced
by a non-adiabatic pressure in the curvaton and the 
inhomogeneous scenarios.

\item We present the exact expression for the
second-order temperature fluctuations on super-horizon scales, extending
the well-known expression for the Sachs-Wolfe effect at linear order.

\item We provide the exact definition for the non-linear parameter 
$f_{\rm NL}$ entering in the determination of higher-order statistics as 
the bispectrum of the temperature anisotropies.

\item We compute the exact expressions for the non-linear parameter
$f_{\rm NL}$ in the different scenarios for the
creation of cosmological perturbations. 

\end{itemize}
The paper is organized as follows. In Section II
we study the time-evolution for the system composed by a generic scalar field
decaying into  radiation which can be used to study a reheating phase.
In Section III we describe in a gauge-invariant manner the 
evolution of cosmological perturbations at linear order while in Section IV
we extend this analysis to second-order providing the evolution equations
for the curvature perturbations. In Section V we apply our findings to
the various scenarios for the generation of the cosmological
perturbations and give the exact expressions for the 
second-order total curvature perturbation. In Section VI we compute
the expression for the second-order temperature fluctuations on large
scales and the corresponding non-linear parameter. In Section VII
the predictions of the non-linear parameter for the various scenarios
are provided. Finally, our conclusions are contained in 
Section VIII.


\section{The metric and the background equations}
In this section we provide the equations describing the 
time-evolution for the system composed by a generic scalar field
and radiation. In particular, 
we will focus on the evolution of cosmological 
perturbations on large scales, up to second-order, for a system composed by 
a scalar field oscillating around the minimum of its potential 
and a radiation fluid having in mind
the physical case of reheating. Averaged over several oscillations the 
effective equation of state of the scalar field $\varphi$ 
is $w_\varphi=\langle P_\varphi/\rho_\varphi 
\rangle = 0$, where $P_\varphi$ and $\rho_\varphi$  are  
the scalar field pressure and energy density respectively. The scalar field 
is thus equivalent to a fluid of non-relativistic particles \cite{Turner}. 
Moreover it is supposed to decay into radiation
(light particles) with a decay rate $\Gamma$. We can thus describe this 
system as a pressureless and a radiation fluid which interact through
an energy transfer triggered by the decay rate $\Gamma$. We follow 
the gauge-invariant approach developed in Ref.~\cite{wmu} to study 
cosmological perturbations at first-order for the general 
case of an arbitrary number of interacting fluids and we shall 
extend the analysis to second-order in the perturbations. 
Indeed the system under study
encompasses the dynamics of the three main mechanisms for the generation of 
the primordial cosmological density perturbations on large scales, namely 
the standard scenario of single field inflation \cite{guth81,lrreview}, 
the curvaton scenario \cite{curvaton1,LW,curvaton3,LUW}, and 
the recently introduced scenario of ``inhomogeneous reheating'' 
\cite{gamma1,gamma2,gamma3,MatRio}.
In particular our second-order results allow us to calculate the level of 
non-Gaussianity in the gravitational potential in the latter 
scenario, where the inflaton decay rate has spatial fluctuations 
$\delta \Gamma$, produced when the couplings of the inflaton field to 
normal matter depends on the vacuum expectations value of some other light 
scalar fields $\chi$ present during inflation.

Let us consider the system composed by the oscillating scalar field
$\varphi$
and the radiation fluid. Each component has energy-momentum tensor 
$T^{\mu\nu}_{(\varphi)}$ and $T^{\mu\nu}_{(\gamma)}$. The total energy 
momentum $T^{\mu\nu}=T^{\mu\nu}_{(\varphi)}+T^{\mu\nu}_{(\gamma)}$ is 
covariantly conserved, but allowing for an interaction between the two fluids
\cite{KSa}
\begin{eqnarray}
 \label{Qvector}
\nabla_\mu T^{\mu\nu}_{(\varphi)}&=&Q^\nu_{(\varphi)}\,, \nonumber\\
\nabla_\mu T^{\mu\nu}_{(\gamma)}&=&Q^\nu_{(\gamma)}\,, 
\end{eqnarray}
where $Q^\nu_{(\varphi)}$ and $Q^\nu_{(\gamma)}$ are
 the generic energy-momentum transfer to
the scalar field and radiation sector respectively
and are  subject to the constraint
\begin{equation}
\label{Qconstraint}
Q^\nu_{(\varphi)}+Q^\nu_{(\gamma)}=0 \,.
\end{equation} 
The energy-momentum transfer $Q^\nu_{(\varphi)}$ and 
$Q^\nu_{(\gamma)}$ can be decomposed for convenience as  \cite{KSa}
\bea
\label{splitQ}
Q^\nu_{(\varphi)}&=&\hat{Q}_{\varphi}u^{\nu}+f_{(\varphi)}^{\nu}
\, , \nonumber \\
Q^\nu_{(\gamma)}&=&\hat{Q}_{\gamma}u^{\nu}+f_{(\gamma)}^{\nu}\, ,
\eea
where the $f^{\nu}$'s are 
required to be orthogonal to the  the total velocity of the fluid  $u^{\nu}$.
The energy continuity equations for the scalar field and radiation 
can be obtained from $u_{\nu}\nabla_\mu T^{\mu\nu}_{(\varphi)}=
u_{\nu}Q^{\nu}_{(\varphi)}$ and $u_{\nu}\nabla_\mu T^{\mu\nu}_{(\gamma)}=
u_{\nu}Q^{\nu}_{(\gamma)}$ and hence from Eq.~(\ref{splitQ})
\bea
\label{conseq}
u_{\nu}\nabla_\mu T^{\mu\nu}_{(\varphi)}&=& \hat{Q}_{\varphi} \, ,\nonumber \\
u_{\nu}\nabla_\mu T^{\mu\nu}_{(\gamma)}&=& \hat{Q}_{\gamma}\, .
\eea  
In the case of an oscillating scalar field decaying into radiation 
the energy transfer coefficient $\hat{Q}_\varphi$ is given 
by~\cite{Qreh}
\bea 
\label{Qreh}
\hat{Q}_\varphi&=&-\Gamma \rho_\varphi \nonumber \\
\hat{Q}_\gamma&=&\Gamma \rho_\varphi \ , ,
\eea
where $\Gamma$ is the decay rate of the scalar field into radiation.
 
A consistent study of the density perturbations 
requires to consider the metric perturbations around a spatially-flat 
Friedmann-Robertson-Walker (FRW) background as well. 
In the following we have used the expression for the metric 
perturbations contained in Refs.~\cite{MMB,noi}. 
The components of a perturbed spatially flat Robertson-Walker 
metric can be written as
\bea \label{metric1}
g_{00}&=&-a^2(\tau)\left( 1+2 \phi^{(1)}+\phi^{(2)} \right)\, ,\nonumber\\
g_{0i}&=&a^2(\tau)\left( \hat{\omega}_i^{(1)}+\frac{1}{2} 
\hat{\omega}_i^{(2)} \right)
\, ,  \nonumber\\g_{ij}&=&a^2(\tau)\left[
(1 -2 \psi^{(1)} - \psi^{(2)})\delta_{ij}+
\left( \hat{\chi}^{(1)}_{ij}+\frac{1}{2}\hat{\chi}^{(2)}_{ij} \right)\right] 
\,,
\eea
where $a(\tau)$ is the scale factor and $\tau=\int dt/a$ is the 
conformal time.
The standard splitting of the perturbations into scalar, transverse 
({\it i.e} divergence-free) vector parts, and transverse trace-free tensor 
parts with respect to the 3-dimensional space with metric $\delta_{ij}$ 
can be performed in the following way:
\begin{equation}
\hat{\omega}_i^{(r)}=\partial_i\omega^{(r)}+\omega_i^{(r)}\, ,
\end{equation}
\begin{equation}
\hat{\chi}^{(r)}_{ij}=D_{ij}\chi^{(r)}+\partial_i\chi^{(r)}_j
+\partial_j\chi^{(r)}_i
+\chi^{(r)}_{ij}\, ,
\end{equation}
where $r=1,2$ stand for the order of the perturbations, $\omega_i$
and $\chi_i$ are transverse vectors ($\partial^i\omega^{(r)}_i=
\partial^i\chi^{(r)}_i=0$), $\chi^{(r)}_{ij}$ is a symmetric transverse and 
trace-free tensor ($\partial^i\chi^{(r)}_{ij}=0$, $\chi^{i(r)}_{~i}
=0$) and $D_{ij}=\partial_i \partial_j - (1/3) \, \, 
\delta_{ij}\, \partial^k\partial_k$ is a trace-free operator.
Here and in the following latin indices 
are raised and lowered using $\delta^{ij}$ and $\delta_{ij}$, respectively.\\
For our purposes the metric in Eq.~(\ref{metric1}) can be simplified. In fact, 
first-order vector perturbations are zero; 
moreover, the tensor part gives a negligible contribution to second-order
perturbations. Thus, in the following we can neglect  
$\omega^{(1)}_i$, $\chi^{(1)}_{i}$ and $\chi^{(1)}_{ij}$.
However the same is not true for the second-order perturbations. 
In the second-order theory the second-order vector and tensor 
contributions can be generated by the first-order scalar perturbations 
even if they are initially zero \cite{MMB}. 
Thus we have to take them into account and we shall use the metric
\bea \label{metric2}
g_{00}&=&-a^2(\tau)\left( 1+2 \phi^{(1)}+\phi^{(2)} \right)\, ,\nonumber\\
g_{0i}&=&a^2(\tau)\left( \partial_i\omega^{(1)}+\frac{1}{2}\, 
\partial_i\omega^{(2)}+\frac{1}{2}\, \omega_i^{(2)} \right)
\, ,  \nonumber\\g_{ij}&=&a^2(\tau)\left[
\left( 1 -2 \psi^{(1)} - \psi^{(2)} \right)\delta_{ij}+
D_{ij}\left( \chi^{(1)} +\frac{1}{2} \chi^{(2)} \right)\right.\nonumber\\
&+&\left.\frac{1}{2}\left( \partial_i\chi^{(2)}_j
+\partial_j\chi^{(2)}_i
+\chi^{(2)}_{ij}\right)\right]
\,.
\eea 
The controvariant metric tensor is obtained by requiring (up to second-order)
that $g_{\mu\nu}g^{\nu\lambda}=\delta_\mu\, ^\lambda$ and it is given by
\bea \label{cont}
g^{00}&=&-a^{-2}(\tau)\left( 1-2 \phi^{(1)}-\phi^{(2)} +4\left(
\phi^{(1)}\right)^2\right.\nonumber\\
&-&\left.\partial^i\omega^{(1)}\partial_i\omega^{(1)}  
\right)\, ,\nonumber\\
g^{0i}&=&a^{-2}(\tau)\left[ \partial^i\omega^{(1)}+\frac{1}{2}
\left( \partial^i\omega^{(2)}+\omega^{i(2)} \right)\right.\nonumber\\
&+&\left. 2 \left( \psi^{(1)}
-\phi^{(1)} \right) \partial^i\omega^{(1)}-\partial^i\omega^{(1)}
D^i\,_k \chi^{(1)} \right]
\, ,  \nonumber\\
g^{ij}&=&a^{-2}(\tau) \left[
\left( 1+2 \psi^{(1)} +\psi^{(2)}+4 \left( \psi^{(1)} \right)^2 
\right) \delta^{ij}\right.\nonumber\\
&-&\left.
D^{ij}\left( \chi^{(1)} +\frac{1}{2} \chi^{(2)} \right)
- \frac{1}{2}\left( \partial^i\chi^{j(2)}
+\partial^j\chi^{i(2)}
+\chi^{ij(2)} \right)\right.\nonumber\\
&-&\left.\partial^i\omega^{(1)}\partial^j\omega^{(1)}-
4 \psi^{(1)}D^{ij}\chi^{(1)}+D^{ik}\chi^{(1)}D^j_{~k} \chi^{(1)} \right]\,.
\eea


\subsection{Background equations}

The evolution of the FRW background 
universe is governed by the Friedmann constraint

\begin{eqnarray}
\label{Friedmann}
\H^2 &=& \frac{8\pi G}{3}\rho a^2 \,,
\end{eqnarray}
and the continuity equation
\begin{equation}
\label{continuity}
\rho'=-3\H\left( \rho+P\right)\,,
\end{equation}
where a prime denotes differentiation with respect to conformal time,
 $\H\equiv a'/a$ is the Hubble parameter in conformal time, 
and $\rho$ and $P$ are the total energy density and the total
pressure of the system.
The total energy density
and the total pressure are related to the energy density and
pressure of the scalar field and radiation  by

\begin{eqnarray}
\rho &=&\rho_\varphi+\rho_\gamma \,, \nonumber\\
P&=& P_\varphi+P_\gamma \,, 
\end{eqnarray}
where $P_\gamma$ is the radiation pressure. The energy continuity equations 
for the energy density of the scalar field $\rho_\varphi$ and radiation 
$\rho_\gamma$ in the background are 
\begin{eqnarray}
\label{rhophi'}
\rho_{\varphi}'
&=&-3\H\left(\rho_{\phi}+P_{\phi}\right) +a Q_{\varphi}\, , \\
\label{rhog'}
\rho_{\gamma}'
&=&-4\H\left(\rho_{\gamma}+P_{\gamma}\right) +a Q_{\gamma}\, ,
\label{m}
\end{eqnarray}
where $Q_{\varphi}$ and $Q_{\gamma}$ indicate the background values of 
the transfer coefficient $\hat{Q}_{\varphi}$ and $\hat{Q}_{\gamma}$, 
respectively.


\section{Gauge-invariant perturbations at first-order}
The primordial adiabatic density perturbation is associated with a 
perturbation in the spatial curvature $\psi$ and it is usually characterized 
in a gauge-invariant manner by the curvature perturbation on 
hypersurfaces of uniform total density $\rho$, usually indicated with 
$\zeta$. Indeed, 
both the density perturbations, $\delta\rho$, $\delta\rho_\phi$ and $
\delta\rho_\gamma$, and the
curvature perturbation, $\psi$, are in general gauge-dependent.
Specifically they depend upon the chosen time-slicing in an
inhomogeneous universe. The curvature perturbation on fixed time hypersurfaces
is a gauge-dependent quantity: after an arbitrary linear coordinate
transformation, $\tau\rightarrow \tau+\delta \tau$, it transforms
at first-order as
$\psi^{(1)}\rightarrow \psi^{(1)}
+\H\delta \tau$. For a scalar quantity, such as the 
energy density, the corresponding transformation is, at
first-order,  $\delta^{(1)}\rho\rightarrow
\delta^{(1)}\rho-\rho'\delta \tau$. 
However a gauge-invariant combination $\zeta$ can
be constructed which describes the density perturbation on uniform
curvature slices or, equivalently the curvature of uniform density
slices \cite{bardeen83}
\be
\label{zetatot}
\zeta^{(1)}= -\psi^{(1)} - \H\frac{\delta^{(1)}\rho}{\rho'}
\, .
\label{eqzeta}
\ee     
Similarly it is possible to define the curvature perturbations 
$\zeta_{i}$ associated with each individual energy density components, 
which to linear order are defined as \cite{wmu}
\begin{eqnarray}
\label{zeta1phi}
\zeta^{(1)}_{\varphi}
 &=& - \psi^{(1)} - \H\left(\frac{\delta^{(1)}\rho_{\varphi}}
{\rho_{\varphi}'}\right)\, , \\
\label{zeta1g}
\zeta^{(1)}_{\gamma}
 &=& - \psi^{(1)} - \H\left(\frac{\delta^{(1)}\rho_{\gamma}}
{\rho_{\gamma}'}\right)\, .
\end{eqnarray}   
Notice that the total curvature perturbation $\zeta$ can be expressed as a 
weighted sum of the single curvature perturbations of the scalar field 
and radiation fluid as 
\be
\label{zetatot1}
\zeta^{(1)}= f \zeta^{(1)}_{\varphi}+ (1-f)\zeta^{(1)}_\gamma \, .
\ee
where
\be
f=\frac{\rho_\varphi'}{\rho'} \, ,\quad
1-f=\frac{\rho_\gamma'}{\rho'}\, 
\ee
define the contribution of the scalar field and radiation 
to the total curvature perturbation $\zeta^{(1)}$, respectively. 
On the other hand the relative energy density perturbations can be 
characterized in a gauge-invariant manner through the difference between the 
two curvature perturbations 
\be
\label{defS}
{\cal S}^{(1)}_{\phi\gamma}=3(\zeta^{(1)}_\phi-\zeta^{(1)}_\gamma)= -3\H
\left(\frac{\delta^{(1)}\rho_\phi}{\rho_\phi'} 
- \frac{\delta^{(1)}\rho_\gamma}{\rho_\gamma'} \right) \, .
\ee
The quantity ${\cal S}^{(1)}_{\phi\gamma}$ is usually called the entropy or 
isocurvature perturbation \cite{KSa}.

\subsection{Evolution of first-order curvature perturbations on large scales}
The equations of motion for the curvature perturbations $\zeta^{(1)}_\varphi$
and  $\zeta^{(1)}_\gamma$ can be obtained perturbing at first order the 
continuity energy 
equations~(\ref{conseq}) for the scalar field and  
radiation energy densities, 
including the energy transfer. Expanding the transfer coefficients 
$\hat{Q}_{\varphi}$ and $\hat{Q}_{\gamma}$ up to first 
order in the perturbations around the 
homogeneous background as 
\begin{eqnarray}
\hat{Q}_{\varphi}&=& Q_{\varphi}+\delta^{(1)}Q_{\varphi}\, , \\
\hat{Q}_{\gamma}&=& Q_{\gamma}+\delta^{(1)}Q_{\gamma }\, ,
\end{eqnarray}
Eqs.~(\ref{conseq}) give -- on wavelengths larger than the 
horizon scale -- \footnote{Here and in the following , as in Refs.
\cite{mw,BMR2}, we neglect gradient terms which, upon integration over time, 
may give rise to non-local operators. However, note that these gradient terms 
will not affect the gravitational potential bispectrum on large scales.}

\begin{eqnarray} 
\label{pertenergyexact1}
&&{\delta^{(1)}\rho'}_{\varphi}+3\H\left( \delta^{(1)}\rho_{\varphi}
+\delta^{(1)}P_{\varphi} \right)
- 3 \left( \rho_{\varphi}+P_{\varphi} \right)\psi^{(1)'}\nonumber \\ 
&&= a \, Q_{\varphi}\phi^{(1)}+a\,  \delta^{(1)} Q_{\varphi}\, , \\
\label{pertenergyexact2}
&&{\delta^{(1)}\rho'}_{\gamma}+3\H\left( 
\delta^{(1)}\rho_{\gamma} +\delta^{(1)}P_{\gamma} \right)
- 3 \left( \rho_{\gamma}+P_{\gamma} \right) \psi^{(1)'}\nonumber \\
&&= a\, Q_{\gamma}\phi^{(1)}+a\, \delta^{(1)}Q_{\gamma}\, .
\end{eqnarray}
Notice that the oscillating scalar field and radiation have fixed equations 
of state with $\delta^{(1)}P_\varphi = 0$ and $\delta^{(1)}P_\gamma =  
\delta^{(1)}\rho_\gamma/3$ (which correspond to vanishing intrinsic 
non-adiabatic pressure perturbations).
Using the perturbed $(0-0)$-component
of Einstein's equations for super-horizon wavelengths
$\psi^{(1)'}+\H \phi^{(1)}=-\frac{\H}{2}\frac{\delta^{(1)}\rho}{\rho}$, 
we can 
rewrite Eqs. (\ref{pertenergyexact1}) and~(\ref{pertenergyexact2})
in terms of the gauge-invariant curvature
perturbations $\zeta^{(1)}_\varphi$ and $\zeta^{(1)}_\gamma$   \cite{wmu}
\begin{eqnarray}
\label{eq:zeta1phi}
\zeta^{(1)'}_\varphi&=& 
\frac{a\H}{\rho_\varphi'}\bigg[\delta^{(1)}Q_\varphi-
\frac{Q_\varphi'}{\rho_\varphi'} \delta^{(1)} \rho_\varphi\nonumber\\
&+&Q_\varphi \frac{\rho'}{2\rho} 
\bigg( \frac{\delta^{(1)}\rho_\varphi}{\rho_\varphi'} - 
\frac{\delta^{(1)}\rho}{\rho'} \bigg) \bigg]\, ,\\
\label{eq:zeta1g}  
\zeta^{(1)'}_\gamma&=& 
\frac{a\H}{\rho_\gamma'}\left[\delta^{(1)}Q_\gamma-
\frac{Q_\gamma'}{\rho_\gamma'} \delta^{(1)} \rho_\gamma\right.\nonumber\\ 
&+&\left.Q_\gamma \frac{\rho'}{2\rho} 
\left(\frac{\delta^{(1)}\rho_\gamma}{\rho_\gamma'} -
\frac{\delta^{(1)}\rho}{\rho'} \right) \right] \, ,
\end{eqnarray}
where $\delta^{(1)}Q_{\gamma}=-\delta^{(1)}Q_{\varphi}$ from the constraint
in Eq~(\ref{Qconstraint}).
If the energy transfer coefficients $\hat{Q}_\varphi$ and 
$\hat{Q}_\gamma$ are given in terms 
of the decay rate $\Gamma$ as in Eq.~(\ref{Qreh}), the first order perturbation 
are respectively
\begin{eqnarray}
\label{Qgamma1}
\delta^{(1)}Q_\varphi&=&-\Gamma \delta^{(1)}\rho_\varphi-\delta^{(1)}\Gamma 
\, \rho_\varphi \, , \\
\label{Qgamma2}
\delta^{(1)}Q_\gamma &=&\Gamma \delta^{(1)}\rho_\varphi+\delta^{(1)}\Gamma 
\, \rho_\varphi \, ,
\end{eqnarray}   
where notice in particular that we have allowed for a perturbation in the 
decay rate $\Gamma$, 
\be
\Gamma(\tau, {\bf x})=\Gamma(\tau)+\delta^{(1)}\Gamma(\tau, {\bf x})\, .
\ee
Such perturbations in the inflaton decay rate are indeed the key feature 
of the 
``inhomogeneous reheating'' scenario \cite{gamma1,gamma2,gamma3}.
In fact from now on we shall consider the background value $\Gamma$ of the 
decay rate as constant in time, 
$\Gamma \approx \Gamma_{*}$ as this is the case for the
the standard case of inflation and the inhomogeneous reheating mechanism.
In such a case $\delta^{(1)}\Gamma$ is automatically gauge-invariant
(for a gauge-invariant generalization in the case of $\Gamma'\neq 0$, see
Ref. \cite{MatRio}).
Plugging the expressions ~(\ref{Qgamma1}-\ref{Qgamma2}) 
into Eqs.~(\ref{eq:zeta1phi}-\ref{eq:zeta1g}), and using 
Eq.~(\ref{zetatot1}), the first order 
curvature perturbations for the scalar field and radiation obey on large 
scales 
\begin{eqnarray}
\label{zeta1phi'}
\zeta^{(1)'}_\varphi&=&\frac{a \Gamma}{2} \frac{\rho_\varphi}{\rho_\varphi'} 
\frac{\rho'}{\rho} \left( \zeta^{(1)} -\zeta^{(1)}_\varphi 
\right)+a \H \frac{\rho_\varphi}{\rho_\varphi'} \delta^{(1)} \Gamma \, ,\\
\label{zeta1g'}
\zeta^{(1)'}_\gamma&=&-\frac{a}{\rho_\gamma'} \left[ \Gamma \rho' 
\frac{\rho_\varphi'}{\rho_\gamma'}\left(1-\frac{\rho_\varphi}{2\rho} 
\right)\left( \zeta^{(1)} -\zeta^{(1)}_\varphi 
\right)
+\H\rho_\varphi \delta^{(1)}\Gamma 
\right]\, . \nonumber \\
&&
\end{eqnarray}
From Eq.~(\ref{zetatot1}) it is thus possible to find the equation of 
motion for the total curvature perturbation $\zeta^{(1)}$ using the evolution 
of the individual curvature perturbations in Eqs.~(\ref{zeta1phi'}) 
and~(\ref{zeta1g'})
\begin{eqnarray}
\label{zeta1'}
\zeta^{(1)'}&=&f'\left( \zeta^{(1)}_\varphi- \zeta^{(1)}_\gamma \right)
+f \zeta^{(1)'}_\varphi+(1-f)\zeta^{(1)'}_\gamma \nonumber \\
&=& \H f (1-f)\left( \zeta^{(1)}_\varphi-\zeta^{(1)}_\gamma \right)=
-\H f \left( \zeta^{(1)}-\zeta^{(1)}_\varphi \right)\, .
\end{eqnarray}
Notice that during the decay of the scalar field into the radiation fluid, 
$\rho_{\gamma}'$ in Eq.~(\ref{zeta1g'}) may vanish. 
So it is convenient to close the system of equations by using 
the two equations~(\ref{zeta1phi'}) and~(\ref{zeta1'}) for the evolution of 
$\zeta^{(1)}_{\varphi}$ and $\zeta^{(1)}$.


\section{Second-order curvature perturbations}
We now generalize to second-order in the density perturbations
the results of the previous section. In particular we obtain an equation of 
motion on large scales for the individual second-order curvature 
perturbations which include also the energy transfer between the scalar field 
and the radiation component.

As it has been shown in Ref.~\cite{mw} it is possible to define the 
second-order curvature  perturbation on uniform total density 
hypersurfaces by the quantity  (up to a gradient term) 
\bea
\label{qqq}
\zeta^{(2)}&=&-\psi^{(2)}-\H\frac{\delta^{(2)}\rho}{\rho^\prime}\nonumber \\
&+&2\H\frac{\delta^{(1)}\rho^\prime}{\rho^\prime}
\frac{\delta^{(1)}\rho}{\rho^\prime}
+2\frac{\delta^{(1)}\rho}{\rho^\prime}\left(\psi^{(1)\prime}
+ 2\H\psi^{(1)}\right) \nonumber \\
&-&\left(\frac{\delta^{(1)}\rho}{\rho^\prime}
\right)^2 \left(\H \frac{\rho^{\prime\prime}}{\rho^\prime}-
\H^\prime-2\H^2\right) \, ,
\eea
where the curvature perturbation $\psi$ has been expanded up to 
second-order as $\psi=\psi^{(1)}+\frac{1}{2} \psi^{(2)}$ and 
$\delta^{(2)} \rho$ 
corresponds to the second-order perturbation in the total 
energy density around the homogeneous background  $\rho(\tau)$ 
\bea
\label{rho}
\rho(\tau, \bfx)&=&\rho(\tau)+\delta\rho(\tau, \bfx)\nonumber \\
&=&\rho(\tau)+
\delta^{(1)}\rho(\tau, \bfx)+\frac{1}{2}\delta^{(2)}
\rho(\tau, \bfx)\, .
\eea
The quantity $\zeta^{(2)}$ is gauge-invariant and, as its first-order 
counterpart defined in Eq.~(\ref{zetatot}),  
it is sourced on superhorizon scales 
by a second-order non-adiabatic pressure perturbation \cite{mw}. 

In the standard scenario
where the generation of cosmological perturbations is induced by 
fluctuations of a single inflaton field (and there is no curvaton) 
the evolution of the perturbations is purely adiabatic, and the total 
curvature perturbation $\zeta^{(2)}$ is indeed conserved. 
In Ref.~\cite{BMR} 
the conserved quantity $\zeta^{(2)}$ has been used     
to follow the evolution  on large scales of 
the primordial non-linearity in the cosmological perturbations from a period
of  
inflation to the matter dominated era. On the contrary in 
the curvaton and inhomogeneous reheating scenarios the total 
curvature perturbation $\zeta^{(2)}$
evolves on large scales due to a non-adiabatic pressure. 
In a similar manner to the linear order it is possible to follow the evolution 
of $\zeta^{(2)}$ through the evolution of the density perturbations in the 
scalar field and radiation.

Let us introduce now the 
curvature perturbations $\zeta^{(2)}_i$ at second-order for each 
individual component. Such quantities will be given by the same formula 
as Eq.~(\ref{qqq}) relatively to each energy density $\rho_i$ 
\bea
\label{zeta2singole}
\zeta^{(2)}_i&=&-\psi^{(2)}-\H\frac{\delta^{(2)}\rho_i}
{\rho_{i}^\prime}\nonumber \\
&+&2\H\frac{\delta^{(1)}\rho_{i}^\prime}{\rho_{i}^\prime}
\frac{\delta^{(1)}\rho_i}{\rho_{i}^\prime}
+2\frac{\delta^{(1)}\rho_i}{\rho_{i}^\prime}\left(\psi^{(1)\prime}
+ 2\H\psi^{(1)}\right) \nonumber \\
&-&\left(\frac{\delta^{(1)}\rho_i}{\rho_{i}^\prime}
\right)^2 \left(\H \frac{\rho_{i}^{\prime\prime}}{\rho_{i}^\prime}-
\H^\prime-2\H^2\right) \, .
\eea   
Since the quantities $\zeta^{(1)}_i$ and $\zeta^{(2)}_i$  
are gauge-invariant, we choose to 
work in the spatially flat gauge $\psi^{(1)}=\psi^{(2)}=0$ 
if not otherwise specified.
Note that from Eqs.~(\ref{zeta1phi}) and (\ref{zeta1g})
$\zeta^{(1)}_\varphi$ and $\zeta^{(1)}_\gamma$
are thus given by
\begin{eqnarray}
\label{zeta1phiflat}
\zeta^{(1)}_{\varphi}
 &=& - \H\left(\frac{\delta^{(1)}\rho_{\varphi}}
{\rho_{\varphi}'}\right)\, , \\
\label{zeta1gflat}
\zeta^{(1)}_{\gamma}
 &=& - \H\left(\frac{\delta^{(1)}\rho_{\gamma}}
{\rho_{\gamma}'}\right)\, .
\end{eqnarray} 
Using Eqs.~(\ref{zeta1phiflat}) and ~(\ref{zeta1gflat}),
the energy continuity equations at first order ~(\ref{pertenergyexact1}) 
and~(\ref{pertenergyexact2}) in the spatially flat gauge $\psi^{(1)}=0$ yield 
\bea
\frac{\delta^{(1)} \rho'}{\rho'}&=&3f\zeta^{(1)}_\varphi+4(1-f) 
\zeta^{(1)}_\gamma\, ,\nonumber \\
\H \frac{\delta^{(1)} \rho}{\rho'}&=&- f \zeta^{(1)}_\varphi- (1-f) 
\zeta^{(1)}_\gamma\, .
\eea
We can thus rewrite the total second-order 
curvature perturbation $\zeta^{(2)}$ in Eq.~(\ref{qqq}) as
\begin{eqnarray}
\zeta^{(2)}&=&-\H \frac{\delta^{(2)} \rho}{\rho'} \nonumber \\
&-&\left[ f \zeta^{(1)}_\varphi+(1-f) \zeta^{(1)}_\gamma\right] 
\left[ f^2 \zeta^{(1)}_\varphi+(1-f)(2+f) \zeta^{(1)}_\gamma\right] 
\, , \nonumber
\\
&&
\end{eqnarray}
where we have used the background equations~(\ref{rhophi'}) and~(\ref{rhog'})
to find $\H \frac{\rho''}{\rho'}-\H'-2\H^2=-\H^2 (6-f)$.
  
Following the same procedure the individual curvature perturbations 
defined in Eq.~(\ref{zeta2singole}) are given by
\bea
\label{zeta2phiflat}
\zeta^{(2)}_\varphi&=&-\H \frac{\delta^{(2)}\rho_\varphi}{\rho_\varphi'}
+[2-3(1+w_\varphi)]\left( \zeta^{(1)}_\varphi \right)^2\nonumber \\
&-&2
\left( a \frac{Q_\varphi \phi^{(1)}}{\rho_\varphi'}+
a \frac{\delta^{(1)} Q_\varphi}{\rho_\varphi'} \right) \zeta^{(1)}_\varphi
\nonumber \\
&-& 
\left[ a \frac{Q_\varphi'}{\H \rho_\varphi'}-\frac{a}{2} 
\frac{Q_\varphi}{\H \rho_\varphi'} \frac{\rho'}{\rho} \right]
\left( \zeta^{(1)}_\varphi \right)^2 \, , \\
\label{zeta2gflat}
\zeta^{(2)}_\gamma&=&-\H \frac{\delta^{(2)}\rho_\gamma}{\rho_\gamma'}
+[2-3(1+w_\gamma)]\left( \zeta^{(1)}_\gamma \right)^2\nonumber \\
&-&2
\left( a \frac{Q_\gamma \phi^{(1)}}{\rho_\gamma'}+
a \frac{\delta^{(1)} Q_\gamma}{\rho_\gamma'} \right)
\zeta^{(1)}_\gamma
\nonumber \\
&-&  
\left[ a \frac{Q_\gamma'}{\H \rho_\gamma'}-\frac{a}{2} 
\frac{Q_\gamma}{\H \rho_\gamma'} \frac{\rho'}{\rho} \right]
 \left( \zeta^{(1)}_\gamma \right)^2 \, ,
\eea
where $w_\gamma=1/3$ is the radiation equation of state.
Using  Eqs.~(\ref{zeta2phiflat}) and~(\ref{zeta2gflat}) to 
express the perturbation of the total energy density $\delta^{(2)}\rho$
one obtains the following expression for the total curvature 
perturbation
$\zeta^{(2)}$
\bea
\label{main0}
\zeta^{(2)}&=&f \zeta^{(2)}_\varphi+(1-f) \zeta^{(2)}_\gamma\nonumber \\
&+&f(1-f)(1+f)\left( \zeta^{(1)}_\varphi
-\zeta^{(1)}_\gamma\right)^2 \nonumber \\
&+&2
\left( a \frac{Q_\varphi \phi^{(1)}}{\rho'}+
a \frac{\delta^{(1)} Q_\varphi}{\rho'} \right)
\left[\zeta^{(1)}_\varphi- \zeta^{(1)}_{\gamma} \right]\nonumber \\
&+& \left( a \frac{Q_\varphi'}{\H \rho'}-\frac{a}{2} 
\frac{Q_\varphi}{\H\rho} \right)
\left[ \left( \zeta^{(1)}_\varphi \right)^2-
\left( \zeta^{(1)}_\gamma \right)^2 \right]   \nonumber \\
&& 
\eea 
The (0-0)-component of Einstein equations in the spatially flat 
gauge at first-order $\psi^{(1)}=0$
\be
\label{00first}
\phi^{(1)}=-\frac{1}{2} \frac{\delta^{(1)}\rho}{\rho}=\frac{1}{2} 
\frac{\rho'}{\H \rho} \zeta^{(1)}\, ,
\ee
and the explicit expressions of the first order perturbed coefficients 
in terms of the decay rate $\Gamma$, Eqs.~(\ref{Qgamma1}-\ref{Qgamma2}), yield
\bea
\label{main1}
\zeta^{(2)}&=&f \zeta^{(2)}_\varphi+(1-f) \zeta^{(2)}_\gamma
+f(1-f)(1+f)\left( \zeta^{(1)}_\varphi-\zeta^{(1)}_\gamma\right)^2 
\nonumber \\
&+&\frac{a\ \Gamma}{\H} f \left( \zeta^{(1)}_\varphi-
\zeta^{(1)}_\gamma \right)^2 -2 a \delta^{(1)} 
\Gamma \frac{\rho_\varphi}{\rho'} 
\left( \zeta^{(1)}_\varphi-\zeta^{(1)}_\gamma \right) \nonumber \\
&+&\frac{a\Gamma}{\H} (1-2f) \frac{\rho_\varphi}{2\rho}
\left( \zeta^{(1)}_\varphi-\zeta^{(1)}_\gamma \right)^2\, .
\eea 
Eq.~(\ref{main1}) is the generalization to second-order in the perturbations 
of the weighted sum in Eq.~(\ref{zetatot1}) and extends the 
expression 
found in Ref.~\cite{BMR2} in the particular case of 
the curvaton scenario, under the sudden decay approximation, 
where the energy transfer was neglected.

\subsection{Large-scale evolution of second-order curvature perturbations}

In this section we give the equations of motion on large scales 
for the individual second-order curvature perturbations 
$\zeta^{(2)}_\varphi$ and $\zeta^{(2)}_\gamma$. 
The energy transfer coefficients $\hat{Q}_\varphi$ and $\hat{Q}_\gamma$ in 
Eqs.~(\ref{splitQ})
perturbed at second-order around the homogeneous backgrounds are given by
\begin{eqnarray}
\label{Q2phi}
\hat{Q}_{\varphi}&=& Q_{\varphi}+\delta^{(1)}Q_{\varphi}
+\frac{1}{2}\delta^{(2)}Q_\varphi\, , \\
\label{Q2g}
\hat{Q}_{\gamma}&=& Q_{\gamma}+\delta^{(1)}Q_{\gamma }\, ,
+\frac{1}{2}\delta^{(2)}Q_\gamma \, .
\end{eqnarray}
Note that from Eq.~(\ref{Qconstraint}) it follows 
$\delta^{(2)}Q_\gamma=-\delta^{(2)}Q_\gamma$. 
Thus the energy continuity equations~(\ref{conseq}) perturbed at second-order 
give on large scales
\bea
\label{cont2phi}
&&{\delta^{(2)}\rho_\varphi}'+3\H \left(
\delta^{(2)}\rho_\varphi+\delta^{(2)}P_\varphi \right)
-3(\rho_\varphi+P_\varphi) \psi^{(2)'} \nonumber \\ 
&-&6\psi^{(1)'}\left[ \delta^{(1)}\rho_\varphi+\delta^{(1)}P_\varphi+
2 (\rho_\varphi+P_\varphi ) \psi^{(1)}\right] = \nonumber \\
&a& \delta^{(2)}Q_\varphi+a\,  Q_\varphi \phi^{(2)}
-a Q_\varphi \left( \phi^{(1)} \right)^2+2a  \phi^{(1)}\delta^{(1)}Q_\varphi
\, ,
\eea
\bea
\label{cont2g}
&&{\delta^{(2)}\rho_\gamma}'+3\H \left(
\delta^{(2)}\rho_\gamma+\delta^{(2)}P_\gamma \right)
-3(\rho_\gamma+P_\gamma) \psi^{(2)'} \nonumber \\ 
&-&6\psi^{(1)'}\left[ \delta^{(1)}\rho_\gamma+\delta^{(1)}P_\gamma+
2 (\rho_\gamma+P_\gamma ) \psi^{(1)}\right] = \nonumber \\
&a& \delta^{(2)}Q_\gamma+a\, Q_\gamma \phi^{(2)}
-a Q_\gamma \left( \phi^{(1)} \right)^2+2a  
\phi^{(1)}\delta^{(1)}Q_\gamma 
\, ,
\eea
where $\phi^{(2)}$ is the second-order perturbation in the gravitational 
potential $\phi=\phi^{(1)}+\frac{1}{2}\phi^{(2)}$.  
Note that Eqs.~(\ref{cont2phi}) and~(\ref{cont2g}) hold in a generic gauge.
We can now recast such equations in terms of the gauge-invariant 
curvature perturbations  $\zeta^{(2)}_\varphi$ and $\zeta^{(2)}_\gamma$ in 
a straightforward way by choosing the spatially flat gauge $\psi^{(1)}=
\psi^{(2)}=0$. 

The (0-0)-component 
of Einstein equations in the spatially flat gauge at first order     
is given by Eq.~(\ref{00first}), and at second-order on large scales 
it reads (see Eqs.~(A.39) in Ref.~\cite{noi})
\be
\label{00second}
\phi^{(2)}=-\frac{1}{2}\frac{\delta^{(2)}\rho}{\rho}+4\left( \phi^{(1)} \right)^2
\, .
\ee
Using Eqs.~(\ref{00first}) and~(\ref{00second}) 
with the expressions~(\ref{zeta2phiflat}-\ref{zeta2gflat}) 
we find from the energy continuity equations~(\ref{cont2phi}-\ref{cont2g}) 
that the individual second-order curvature perturbations obey on large scales
\bea
\label{eq:zeta2phi}
&&\zeta^{(2)'}_\varphi=\nonumber \\
&-&\frac{a \H}{\rho'}\left[\left( \delta^{(2)}Q_\varphi
-\frac{Q_\varphi'}{\rho_\varphi'}\delta^{(2)}\rho_\varphi \right) 
+Q_\varphi\frac{\rho'}{2\rho}
\left(\frac{\delta^{(2)}\rho_\varphi}{\rho_\varphi'}
-\frac{\delta^{(2)}\rho}{\rho'} \right)
\right] \nonumber \\
&-&3a Q_\varphi \frac{\H}{\rho_\varphi'} \left({\phi^{(1)}}\right)^2
-2a \frac{\H}{\rho_\varphi'}\delta^{(1)}Q_\varphi \phi^{(1)}
-2\zeta^{(1)}_\varphi\zeta^{(1)'}_\varphi\, \nonumber \\
&-&2\left[ \zeta^{(1)}_\varphi
\left( a \frac{Q_\varphi \phi^{(1)}}{\rho_\varphi'}+
a \frac{\delta^{(1)} Q_\varphi}{\rho_\varphi'} \right) \right]^{\prime}
\nonumber \\
&-& \left[ \left(\zeta^{(1)}_\varphi\right)^2 
\left( a \frac{Q_\varphi'}{\H \rho_\varphi'}-\frac{a}{2} 
\frac{Q_\varphi}{\H \rho_\varphi'} \frac{\rho'}{\rho} \right) \right]^{\prime}
\eea
and 
\bea
\label{eq:zeta2g}
&&\zeta^{(2)'}_\gamma=\nonumber \\
&-&\frac{a \H}{\rho'}\left[\left( \delta^{(2)}Q_\gamma
-\frac{Q_\gamma'}{\rho_\gamma'}\delta^{(2)}\rho_\gamma \right) 
+Q_\gamma\frac{\rho'}{2\rho}\left(\frac{\delta^{(2)}\rho_\gamma}{\rho_\gamma'}
-\frac{\delta^{(2)}\rho}{\rho'} \right)
\right] \nonumber \\
&-&3a Q_\gamma \frac{\H}{\rho_\gamma'} \left({\phi^{(1)}}\right)^2
-2a \frac{\H}{\rho_\gamma'}\delta^{(1)}Q_\gamma \phi^{(1)}
-4\zeta^{(1)}_\gamma\zeta^{(1)'}_\gamma\nonumber \\
&-&2\left[ \zeta^{(1)}_\gamma
\left( a \frac{Q_\gamma \phi^{(1)}}{\rho_\gamma'}+
a \frac{\delta^{(1)} Q_\gamma}{\rho_\gamma'} \right) \right]^{\prime}
\nonumber \\
&-& \left[ \left(\zeta^{(1)}_\gamma\right)^2 
\left( a \frac{Q_\gamma'}{\H \rho_\gamma'}-\frac{a}{2} 
\frac{Q_\gamma}{\H \rho_\gamma'} \frac{\rho'}{\rho} \right) \right]^{\prime}\,
\eea
where we have used that $w_{\varphi}=0$ and $w_\gamma=1/3$.

Equations~(\ref{eq:zeta2phi}) and~(\ref{eq:zeta2g}) 
allow to follow the time-evolution of the gauge-invariant
curvature perturbation at second-order.


\section{Evolution of second-order cosmological perturbations in the various
scenarios}
The results contained in the previous section can be used in order to study 
the evolution of the second-order curvature perturbations during the 
reheating phase after a period
of standard single field inflation, and 
in the alternative scenarios for the generation of the primordial adiabatic 
perturbations which have been recently proposed, namely 
the curvaton scenario \cite{curvaton1,LW,curvaton3} 
and the inhomogeneous reheating \cite{gamma1,gamma2,gamma3}.
In fact in each of these scenarios a scalar field is oscillating around the 
minimum of its potential and eventually it decays into radiation.
The evolution at second-order of the curvature perturbations is 
necessary in order to follow the non-linearity of the
cosmological perturbations and thus to accurately compute the level of 
non-Gaussianity  
including all the relevant second-order effects.    
We shall first extend some of the results previously obtained in 
Ref.~\cite{BMR2} for the standard scenario  and the curvaton scenario.
Then we will  present the calculation of  the second-order 
total curvature perturbation produced in the ``inhomogeneous reheating'' 
by the spatial fluctuations of the inflaton decay rate $\Gamma$.

\subsection{Standard Scenario} 
In the standard inflationary scenario 
the observed density perturbations are due to the 
fluctuations of the inflaton field itself. When inflation ends, 
the inflaton oscillates about the minimum of its
potential and decays, thereby reheating the universe. 
In such a standard scenario the inflaton decay rate has no spatial 
fluctuations. 
Here we want to show that in fact during the reheating phase 
the curvature perturbations $\zeta^{(1)}$ and 
$\zeta^{(2)}$ do remain constant on superhorizon scales. 
Eq.~(\ref{zeta1'}) and Eq.~(\ref{zeta1phi'}) with $\delta^{(1)} 
\Gamma=0 $ now read
\bea
\label{zeta1'g}
\zeta^{(1)'}&=& -\H f \left( \zeta^{(1)}-\zeta^{(1)}_\varphi \right)\, , \\
\label{zeta1phi'g}
\zeta^{(1)'}_\varphi&=&\frac{a \Gamma}{2} \frac{\rho_\varphi}{\rho_\varphi'} 
\frac{\rho'}{\rho} \left( \zeta^{(1)} -\zeta^{(1)}_\varphi 
\right)\, .
\eea
At the beginning of the reheating phase, after the end of inflation, 
the total curvature perturbation is initially given by 
the curvature perturbation of the inflaton fluctuations  
$\zeta^{(1)}_{\rm in}=\zeta^{(1)}_{\varphi,{\rm in}}$. Therefore 
Eqs.~(\ref{zeta1'g}) and (\ref{zeta1phi'g}) show that during the 
reheating phase $\zeta^{(1)}=\zeta^{(1)}_\varphi=
\zeta^{(1)}_{\varphi,{\rm in}}$ is a fixed-point of the time-evolution.
This result has been shown at first-order in \cite{MatRio} and extended to 
second-order in the perturbations  in Ref.~\cite{BMR2} under the sudden 
decay approximation. Under such an approximation the individual energy density 
perturbations (and hence the corresponding curvature perturbations) are 
separately conserved until the decay of the scalar field, which amounts 
to saying that in the equations for the curvature perturbations 
Eqs.~(\ref{eq:zeta2phi}) and~(\ref{eq:zeta2g}) one can drop the energy 
transfer triggered by the decay rate $a \Gamma/\H \ll 1$. 
Going beyond the sudden decay approximation, the first order 
results $\zeta^{(1)}_\varphi=\zeta^{(1)}$ in Eq.~(\ref{main1}) yield
\be
\label{zeta2reheating}
\zeta^{(2)}=f \zeta^{(2)}_\varphi+(1-f) \zeta^{(2)}_\gamma\, .
\ee
Deriving this expression and using 
Eqs.~(\ref{eq:zeta2phi}) and~(\ref{eq:zeta2g}) with 
$\delta^{(2)} \Gamma=0$ and  $\zeta^{(1)}_\varphi=\zeta^{(1)}$ the 
equation of motion for $\zeta^{(2)}$ on large scales reads
\be
\label{zeta2'reheating}
\zeta^{(2)'}= -\H f \left( \zeta^{(2)}-\zeta^{(2)}_\varphi \right)\, .
\ee
In the same way as at first order
from Eqs.~(\ref{zeta2reheating}) and~(\ref{zeta2'reheating}) it follows 
that the 
second-order curvature perturbation $\zeta^{(2)}$ remains constant on large 
scales during 
the reheating phase, 
being given at the end of inflation by the curvature perturbation in the 
inflaton field $\zeta^{(2)}_{\rm in}=\zeta^{(2)}_{\varphi, {\rm in}}$.

\subsection{Curvaton scenario}
The evolution of cosmological perturbations in the curvaton scenario 
 has been studied at  second-order in Ref. \cite{BMR2} and here we only 
briefly summarize the main results. 
In the curvaton scenario  the final curvature perturbations
are produced from an initial isocurvature perturbation associated to the
quantum fluctuations of a light scalar field (other than the inflaton), 
the curvaton $\sigma$, 
whose energy density is negligible during inflation. The 
curvaton isocurvature perturbations are transformed into adiabatic
ones when the curvaton decays into radiation much after the end 
of inflation \cite{LW,LUW}. During inflation, since no curvature
perturbation is produced $\zeta^{(1)}=\zeta^{(2)}=0$. After the
end of inflation, the curvaton starts to oscillate when its mass
is of the order of the Hubble rate and the first-order
total curvature perturbation
can be expressed as

\be
\label{zetasum}
\zeta^{(1)}=f\zeta^{(1)}_\sigma+(1-f)\zeta^{(1)}_\gamma \, ,
\ee
where $f=\rho_\sigma'/\rho'$ and we have set $\zeta^{(1)}_\gamma=0$
consistently with the curvaton hypothesis.
When the curvaton energy density is subdominant, the 
density perturbation in the curvaton field $\zeta^{(1)}_\sigma$ gives a 
negligible contribution to the total curvature perturbation, 
thus corresponding to an isocurvature (or entropy) perturbation. 
On the other hand during the oscillations $\rho_\sigma \propto a^{-3}$ 
increases with respect to the energy density of radiation 
$\rho_\gamma \propto a^{-4}$, and the perturbations in the curvaton field 
are then converted into the curvature perturbation.     
After the decay of the curvaton, during the conventional radiation and 
matter dominated  eras the total curvature perturbation  
will remain constant on superhorizon scales at a value which, 
in the sudden decay approximation, is given by  

\be
\label{atcurvdecay}
\zeta^{(1)}=f_D\, \zeta^{(1)}_\sigma \, ,
\ee
where $D$ stands for the epoch of the curvaton decay.
Going beyond the sudden decay approximation it is possible to introduce a 
transfer parameter $r$ defined as \cite{LUW,wmu}
\be
\zeta^{(1)}=r\zeta^{(1)}_\sigma \, ,
\ee  
where $\zeta^{(1)}$ is evaluated well after the epoch of the curvaton 
decay and $\zeta^{(1)}_\sigma$ is evaluated well before this epoch.
The numerical study of the coupled perturbation equations has been performed 
in Ref.~\cite{wmu} showing that the sudden decay approximation is exact when 
the curvaton dominates the energy density before it decays $(r=1)$, while 
in the opposite case   
\be
r\approx \left( \frac{\rho_\sigma}{\rho} \right)_D.
\ee
At second-order, using Eqs. (\ref{eq:zeta2phi}) and (\ref{eq:zeta2g})
under the sudden-decay approximation, the individual curvature
perturbations $\zeta^{(2)}_\sigma$ and $\zeta^{(2)}_\gamma$
are separately conserved and 
the total curvature perturbation 
$\zeta^{(2)}$ reads \cite{BMR2}
\bea
\label{main}
\zeta^{(2)}&=&f \zeta^{(2)}_\sigma+(1-f) \zeta^{(2)}_\gamma\nonumber \\
&+&f(1-f)(1+f)\left( \zeta^{(1)}_\sigma-\zeta^{(1)}_\gamma\right)^2\, .
\eea
The second-order curvature perturbation in the standard 
radiation or matter eras will remain constant on superhorizon scales and, 
in the sudden decay approximation, it is thus given by 
the quantity in Eq.~(\ref{main}) evaluated at the epoch of the curvaton 
decay
\be
\label{zeta2decay}
\zeta^{(2)}=r \zeta^{(2)}_\sigma+r \left( 1-r^2 \right) \left( 
\zeta^{(1)}_\sigma \right)^2 \, ,
\ee 
where we have used the curvaton hypothesis that the curvature perturbation
in the radiation produced at the end of inflation is negligible so that 
$\zeta^{(1)}_\gamma\approx 0$ and 
$\zeta^{(2)}_\gamma\approx 0$. Taking into account that
$\zeta_\sigma^{(2)}=\frac{1}{2}\left(\zeta_\sigma^{(1)}\right)^2$ \cite{BMR2},
one finally obtains
the 
curvature perturbation during the standard radiation or matter dominated 
eras 
\be
\label{zetafinal}
\zeta^{(2)}=r \left( \frac{3}{2}-r^2 \right) 
\left( \zeta^{(1)}_\sigma \right)^2\, .
\ee

\subsection{Inhomogeneous reheating Scenario: $\delta \Gamma \neq 0$}          

Recently, another mechanism for the generation of cosmological
perturbations has been proposed \cite{gamma1,gamma2,gamma3}.  
It acts during the reheating
stage after inflation and it was dubbed the ``inhomogeneous reheating'' 
mechanism in Ref. \cite{gamma3}. The coupling of the inflaton to normal matter 
may be  determined by the vacuum expectation value of fields $\chi$'s
of the underlying theory. 
If those fields are light during inflation, 
fluctuations  $\delta\chi\sim\frac{H}{2\pi}$,
where $H$ is the Hubble rate during inflation, are left imprinted
on super-horizon scales. These perturbations
lead to spatial 
fluctuations in the decay rate $\Gamma$ 
of the inflaton field to ordinary 
matter, $\frac{\delta\Gamma}{\Gamma}
\sim\frac{\delta\chi}{\chi}$,
causing  adiabatic perturbations
in  the final reheating temperature
in different regions of the universe. 
Using cosmic time $dt= a\, d \tau$ 
the first order equation~(\ref{zeta1phi'}) for $\zeta^{(1)}_\varphi$ 
on large scales now reads
\be
\label{zeta1phidot}
\dot{\zeta}^{(1)}_\varphi
=\frac{\Gamma}{2} \frac{\rho_\varphi}{\dot{\rho_\varphi}} 
\frac{\dot{\rho}}{\rho} \left( \zeta^{(1)} -\zeta^{(1)}_\varphi 
\right)+H \frac{\rho_\varphi}{\dot{\rho_\varphi}} \delta^{(1)} \Gamma\, ,
\ee 
where $H=\dot{a}/{a}$ is the Hubble parameter in cosmic time and the dots
stand for differentation with respect to cosmic time.
Perturbing the energy transfer $\hat{Q}_\varphi=-\Gamma \rho_\varphi$ up to 
second-order and expanding the decay rate as 
\be
\label{Gamma2}
\Gamma=\Gamma_{*}+\delta \Gamma=\Gamma_{*}+\delta^{(1)}\Gamma+\frac{1}{2}
\delta^{(2)}\Gamma\, ,
\ee
it follows from Eqs.~(\ref{Q2phi}) and~(\ref{Gamma2})
\be
\label{Q2gamma}
\delta^{(2)}Q_\varphi=-\rho_\varphi \delta^{(2)} \Gamma -
\Gamma_{*} \delta^{(2)}\rho_\varphi -2\delta^{(1)}\Gamma 
\delta^{(1)}\rho_\varphi\, .
\ee
Plugging Eq.~(\ref{Q2gamma}) into Eq.~(\ref{eq:zeta2phi}), 
the equation of motion on large scales for the curvature perturbation 
$\zeta^{(2)}_\varphi$ allowing for  possible fluctuations of the decay rate 
$\delta^{(1)} \Gamma$ and $\delta^{(2)} \Gamma$ turns out to be
\bea
\label{zeta2phidot}
\dot{\zeta}^{(2)}_\varphi&=& 
\frac{H}{\dot{\rho_\varphi}} \left( \delta^{(2)} \Gamma \rho_\varphi
+2\delta^{(1)} \Gamma \delta^{(1)} \rho_\varphi \right)\nonumber\\
&-&\frac{\Gamma_* \rho_\varphi}{2} \frac{\dot{\rho}}{\rho} 
\left(\frac{\delta^{(2)}\rho_\varphi}{\dot{\rho_\varphi}}
-\frac{\delta^{(2)}\rho}{\dot{\rho}} \right)
+ 3\Gamma_* \rho_\varphi \frac{H}{\dot{\rho_\varphi}} {\phi^{(1)}}^2\nonumber\\
&+&
\frac{2H}{\dot{\rho_\varphi}} \left( \delta^{(1)}\Gamma \rho_\varphi 
+\delta^{(1)} \rho_\varphi \Gamma_* \right) \phi^{(1)}-2\zeta^{(1)}_\varphi
\dot{\zeta}^{(1)}_\varphi\nonumber \\
&+& 2\left[ \zeta^{(1)}_\varphi \left(\Gamma_* 
\frac{\rho_\varphi}{\dot{\rho_\varphi}}\phi^{(1)}+\delta^{(1)}\Gamma 
\frac{\rho_\varphi}{\dot{\rho_\varphi}} + 
\Gamma_* \frac{\delta^{(1)} \rho_\varphi}{\dot{\rho}_\varphi} 
\right) \right]^{\cdot}
\nonumber \\
&+& \left[ {\zeta^{(1)}_\varphi}^2 \frac{\Gamma}{H} \left(1-
\frac{\rho_\varphi}{\dot{\rho_\varphi}} \frac{\dot{\rho}}{\rho} \right) 
\right]^{\cdot}\, ,
\eea
where we have used the 
fact that the  decay rate $\Gamma$ in the scenario under consideration 
remains constant. 

We shall now solve Eqs.~(\ref{zeta1phidot}) and~(\ref{zeta2phidot}) under 
a ``mixed-sudden decay approximation''. We shall treat the pressureless scalar 
field fluid and radiation fluid as if they are not interacting until the decay 
of the inflaton when $\Gamma \approx H$. Since at the beginning of the 
reheating phase the energy density in radiation is negligible this means 
that $f =\dot{\rho_\varphi}/\dot{\rho} \approx 1$ 
and there is indeed only a single fluid with, from 
Eq.~(\ref{zetatot1}), $\zeta^{(1)} \approx \zeta^{(1)}_\varphi $ 
and $\zeta^{(1)}_\gamma \approx 0$. In fact under such an approximation we  
can neglect all the terms proportional to the decay rate $\Gamma$, \emph{but} we 
allow for the spatial fluctuations of the decay rate. Thus  
the first order equation~(\ref{zeta1phidot}) reads
\be 
\label{zeta1phiapprox}
\dot{\zeta}^{(1)}_\varphi \simeq -\frac{1}{3} \delta^{(1)} \Gamma\, ,
\ee
where we have used $\dot{\rho_\varphi}=-3H\rho_\varphi$ in the sudden 
decay approximation.
Integration over time yields
\be
\label{solution1}
\zeta^{(1)}_\varphi=-\frac{t}{3} \delta^{(1)}\Gamma =-\frac{2}{9} 
\frac{\delta^{(1)} \Gamma}{H}\simeq \zeta^{(1)}\, , 
\ee   
where we have used that during the oscillations of the 
scalar field which dominates the energy density $H=\frac{2}{3t}$.
The inhomogeneous reheating mechanism produces at linear level 
a gravitational potential which after the 
reheating phase, in the radiation dominated epoch, is given by   
(in the longitudinal gauge) \cite{gamma1}
\be
\label{resultgamma}
\psi^{(1)}=\frac{1}{9} \frac{\delta^{(1)} \Gamma}{\Gamma_*}\, .
\ee
During the radiation dominated epoch the usual relation between the 
Bardeen potential and the curvature perturbation is 
$\psi^{(1)}=-\frac{2}{3} \zeta^{(1)}$, and thus from Eq.~(\ref{solution1})
we can set  the ratio $\Gamma_{*}/H_{D}=\frac{3}{4}$ 
at the time of the inflaton decay in order to reproduce the result 
~(\ref{resultgamma}) of Ref.~\cite{gamma1}. 
Thus from Eq.~(\ref{solution1}) it follows that the value of $\zeta^{(1)}$
is
\be
\label{zetagamma}
\zeta^{(1)} \simeq -\frac{1}{6} \frac{\delta^{(1)} \Gamma}{\Gamma_*}\, .
\ee
We can use this result in order to solve the second-order 
equation~(\ref{zeta2phidot}), which under the sudden decay approximation 
and using Eq.~(\ref{00first}) reads
\be
\label{eq2approx}
\dot{\zeta}^{(2)}_\varphi\simeq -\frac{1}{3}\delta^{(2)} \Gamma
-\zeta^{(1)}_\varphi\delta^{(1)} 
\Gamma -2\left(\zeta^{(1)}_\varphi \dot{\zeta}^{(1)}_\varphi
\right)-\frac{2}{3} \left(  \frac{\delta^{(1)} 
\Gamma}{H}\zeta^{(1)}_\varphi \right)^{\cdot}\, .
\ee 
The fluctuations $\delta \Gamma=\delta^{(1)}\Gamma+\frac{1}{2}\delta^{(2)}
\Gamma$ indeed depends on the underlying physics for the coupling of the 
inflaton field to the other scalar field(s) $\chi$. Let us take for example
$\Gamma(t,{\bf x}) \propto \chi^2(t,{\bf x})$. If the scalar field $\chi$ is
very light, its homogeneous value can be treated as constant 
$\chi(t)\approx \chi_*$ and 
during inflation quantum fluctuations $\delta^{(1)} \chi$ 
around its homogeneous value $\chi_{*}$ are left imprinted on 
super-horizon scales. Therefore non-linear fluctuations $\left( 
\delta^{(1)} \chi \right)^2$ of the decay rate $\Gamma$ are produced as well
\be
\label{expandchi}
\Gamma \propto \chi^2=\chi^2_{*}+2\chi_* \delta^{(1)} \chi+
\left( \delta^{(1)} \chi \right)^2\, .
\ee
From Eqs.~(\ref{Gamma2}) and~(\ref{expandchi}) it follows
\bea
\frac{\delta^{(1)}\Gamma}{\Gamma_*}&=&2\frac{\delta^{(1)}\chi}{\chi_{*}}\, ,
\nonumber \\
\label{gammanonlin}
\frac{\delta^{(2)}\Gamma}{\Gamma_*}&=&2\left( 
\frac{\delta^{(1)}\chi}{\chi_{*}} \right)^2=\frac{1}{2} 
\left(\frac{\delta^{(1)}\Gamma}{\Gamma_*} \right)^2\, .
\eea
Using the first order solution 
$\zeta^{(1)}_\varphi=-\frac{t}{3} \delta^{(1)}\Gamma$ and 
Eq.~(\ref{gammanonlin}) in Eq.~(\ref{eq2approx}),
the evolution of $\zeta^{(2)}_\varphi$ on large scales is
\bea
\dot{\zeta}^{(2)}_\varphi &\simeq& -\frac{1}{6\Gamma_*}
\left( \delta^{(1)} \Gamma \right)^2
+\frac{1}{3}\left( \delta^{(1)} \Gamma \right)^2 t
-2\left(\zeta^{(1)}_\varphi \dot{\zeta}^{(1)}_\varphi
\right) \nonumber \\
&-&\frac{2}{3} \left(  \frac{\delta^{(1)} 
\Gamma}{H}\zeta^{(1)}_\varphi \right)^{\cdot}\, .
\eea 
Integration over time yields
\begin{eqnarray}
\label{zeta2expr}
\zeta^{(2)}_\varphi&=&-\frac{t}{6} \frac{\left( \delta^{(1)} \Gamma \right)^2}
{\Gamma_*}
+\frac{1}{6} \left( \delta^{(1)} \Gamma \right)^2 t^2\nonumber\\
&-&\left({\zeta^{(1)}_\varphi}\right)^2-\frac{2}{3} \zeta^{(1)}_\varphi
\frac{\delta^{(1)}\Gamma}{H}\, .
\end{eqnarray}
At the time of inflaton decay $\Gamma/H_{D}=\frac{3}{4}$, and since 
$H=2/3\,t$, it follows $t_{D}=1/ 2 \Gamma_*$.
Thus $\zeta^{(2)}_\varphi$ 
in Eq.~(\ref{zeta2expr}) evaluated at the time $t_D$ of inflaton decay is  
\be
\zeta^{(2)}_\varphi \simeq -\frac{1}{24}\left(
\frac{\delta^{(1)} \Gamma}{\Gamma_*}\right)^2-{\zeta^{(1)}_\varphi}^2-
\frac{1}{2}\zeta^{(1)}_\varphi \frac{\delta^{(1)} \Gamma}{\Gamma_*}\, .       
\ee 
Using Eq.~(\ref{zetagamma}) we finally find
that the total curvature 
perturbation $\zeta^{(2)}$ in the sudden decay approximation is given by  
\be
\label{risfin}
\zeta^{(2)} \simeq \zeta^{(2)}_\varphi\simeq \frac{1}{2} 
\left( \zeta^{(1)} \right)^2\, .
\ee

\section{Second-order temperature fluctuations on large scales}
The goal of this section is to provide the expression for the
second-order temperature fluctuations on large scales which will allow
the exact definition of the non-linear parameter $f_{\rm NL}$. 
From now on, we will adopt the {\it Poisson
gauge} \cite{poisson} which is defined
by the condition $\omega=\chi=\chi_{i}=0$. Then, one scalar
degree of freedom is eliminated from $g_{0i}$ and one scalar
and two vector degrees of freedom from $g_{ij}$. This gauge generalizes the
so-called longitudinal gauge to include vector and tensor modes
and contains a solenoidal vector $\omega_i^{(2)}$.

The second-order expression for the temperature fluctuation 
field in the Poisson gauge has been obtained in Ref. \cite{molmat}, 
by implementing the general formalism introduced in Ref. \cite{pyne}. 
We are interested here in the large-scale limit of that
expression, which allows to unambiguously define the primordial
non-Gaussian contribution. 
Keeping only the large-scale limit of the linear
and second-order terms in Eqs.(2.27) and (2.28) of Ref. \cite{molmat},
we obtain
\begin{equation}
\label{complete}
\frac{\Delta T}{T} = \phi^{(1)}_{\cal E} + \tau^{(1)}_{\cal E} + 
\frac{1}{2}\left(\phi^{(2)}_{\cal E} + \tau^{(2)}_{\cal E}\right) - 
\frac{1}{2} \left(\phi^{(1)}_{\cal E}\right)^2 + 
\phi^{(1)}_{\cal E} \tau^{(1)}_{\cal E}
\;, 
\end{equation}
where $\phi_{\cal E} = \phi^{(1)}_{\cal E} + 
\frac{1}{2}\phi^{(2)}_{\cal E}$ is the lapse perturbations at emission, 
$\tau_{\cal E} = \tau^{(1)}_{\cal E} + \frac{1}{2} \tau^{(2)}_{\cal E}$ 
is the intrinsic fractional temperature fluctuation at emission, 
$\tau_{\cal E} \equiv \Delta T/T\vert_{\cal E}$. Let us recall that,
at linear order $\phi^{(1)}=\psi^{(1)}$. 
In Eq. (\ref{complete}) we dropped all those 
terms which represent  {\it integrated} contributions such as 
Integrated Sachs-Wolfe, Rees-Sciama  
and more complicated second-order integrated effects \cite{rs}.
Indeed, the non-linear parameter $f_{\rm NL}$ as introduced in \cite{ks,kphd}
singles out the large-scale part of the second-order CMB anisotropies.
One should be able to distinguish the secondary integrated terms from the 
large-scale effects thanks to their specific 
angular scale dependence. For the very same reason, we 
disregarded gravitational-lensing, time-delay, 
Doppler terms and all those second-order effects which are characterized 
by a high-$\ell$ harmonic content. 
We finally dropped contributions at the observer position, 
which only modify the {\it monopole} term.

To obtain the intrinsic anisotropy in the photon temperature, we 
can expand the photon energy-density $\rho_\gamma \propto T^4$ 
up to second-order and write 
\begin{equation}
\tau^{(1)}_{\cal E} = \frac{1}{4} \frac{\delta^{(1)} 
\rho_\gamma}{\rho_{\gamma}}\bigg\vert_{\cal E}=
\frac{1}{3} \frac{\delta^{(1)} 
\rho_m}{\rho_{m}}\bigg\vert_{\cal E}\;, 
\end{equation}
where $\rho_{\gamma}$ is the mean photon energy-density, and 
\begin{equation}
\tau^{(2)}_{\cal E} = \frac{1}{4} \frac{\delta^{(2)} 
\rho_\gamma}{\rho_{\gamma}}\bigg\vert_{\cal E} - 
\frac{3}{16}
\left( \frac{\delta^{(1)}
\rho_\gamma}{\rho_{\gamma}}\bigg\vert_{\cal E}\right)^2
\;.
\end{equation}
Next, we need to relate the photon energy-density fluctuation to the
lapse perturbation, which we can easily do by extending the 
standard adiabaticity condition to second-order. At first-order 
the adiabaticity condition reads $\zeta^{(1)}_m=\zeta^{(1)}_\gamma$
and we obtain 
\begin{equation}
\frac{\delta^{(1)} \rho_\gamma}{\rho_{\gamma}} = \frac{4}{3} 
\frac{\delta^{(1)} \rho_m}{\rho_{m}}\, ,
\end{equation}
where $\rho_m$ is the average energy-density of the matter component.
At second-order the adiabaticity condition 
imposes $\zeta^{(2)}_m=\zeta^{(2)}_\gamma$. 
From Eq. (\ref{zeta2singole}) applied to matter and radiation we
find

\begin{equation}
\frac{\delta^{(2)} \rho_\gamma}{\rho_{\gamma}} = \frac{4}{3} 
\frac{\delta^{(2)} \rho_m}{\rho_{m}}+
\frac{4}{9} 
\left(\frac{\delta^{(1)} \rho_m}{\rho_{m}}\right)^2 \;. 
\end{equation} 
In the large-scale limit, the energy constraint, Eqs. (3.3) and (4.4) of 
Ref. \cite{BMR} in the matter dominated era, yields    
\begin{equation}
\frac{\delta^{(1)} \rho_m}{\rho_{m}} = -2 \psi^{(1)}
\end{equation}
and 
\begin{equation}
\frac{\delta^{(2)} \rho_m}{\rho_{m}} = -2 \phi^{(2)} + 
8 \left( \psi^{(1)} \right)^2 \;.
\end{equation}
We finally obtain the fundamental relation\footnote{
One can easily obtain a gauge-invariant definition
of the  temperature fluctuations  $\frac{\Delta T}{T}$ in terms of
 the second-order gauge-invariant lapse function (on large scales)
\begin{eqnarray}
\phi^{(2)}_{\rm GI}&=&\phi^{(2)}+\omega^{(1)}\left[2\left(
\psi^{(1)'}+2\frac{a'}{a}\psi^{(1)}\right)+\omega^{(1)''}\right.\nonumber\\
&+&\left.5
\frac{a'}{a}\omega^{(1)'}+
\left(\frac{a''}{a}+\left(\frac{a'}{a}
\right)^2\right)\omega^{(1)}\right]\nonumber\\
&+&2\omega^{(1)'}\left(2\psi^{(1)}+\omega^{(1)'}\right)+\alpha^{(2)'}+
\frac{a'}{a}\alpha^{(2)},\nonumber
\end{eqnarray}
where 

\begin{eqnarray}
\alpha^{(2)}&=&\omega^{(2)}+3\omega^{(1)}\omega^{(1)'}\nonumber\\
&+&\nabla^{-2}\partial^i\left[-4\psi^{(1)}\partial_i\omega^{(1)}-
\omega^{(1)'}\partial_i\omega^{(1)}-2 \omega^{(1)}
\partial_i\omega^{(1)'}\right].\nonumber
\end{eqnarray}
Notice that in the Poisson gauge
$\phi^{(2)}_{\rm GI}=\phi^{(2)}$.}

\begin{equation}
\label{deltaT/T}
\frac{\Delta T}{T} = \frac{1}{3}\left[ \psi^{(1)}_{\cal E} + 
\frac{1}{2}\left(\phi^{(2)}_{\cal E} -
\frac{5}{3} \left(\psi^{(1)}_{\cal E} \right)^2 \right) \right]\;.
\end{equation}
From Eq. (\ref{deltaT/T}), it is clear that the expression for the
second-order temperature fluctuations is {\it not} a simple
extension of  the first-order Sachs-Wolfe effect $\frac{\Delta T^{(1)}}{T}
= \frac{1}{3}\psi^{(1)}_{\cal E}$ to second-order since it receives
a correction provided by the term $-\frac{5}{3} 
\left(\psi^{(1)}_{\cal E} \right)^2$. 

If we express the lapse function at second-order as a general convolution 
(see, e.g., Ref. \cite{noi}) 
 
\begin{equation}
\label{lll}
\phi = \phi^{(1)} + \frac{1}{2}\phi^{(2)} = \psi^{(1)}
+ f_{\rm NL}^\phi * \left(\psi^{(1)}\right)^2 \;,
\end{equation}
up to a constant offset, from Eq.~(\ref{deltaT/T}) we can 
define  the {\it true} non-linearity parameter $f_{\rm NL}$
which is the quantity actually measurable by high-resolution CMB
experiments, after properly subtracting instrumental noise, foreground 
contributions and small-scale second-order terms. Therefore, we
find  
\begin{equation}
\label{ppp}
f_{\rm NL} = f_{\rm NL}^\phi - \frac{5}{6} \;.
\end{equation} 
We warn the reader that this is the quantity which enters in the 
determination 
of higher-order statistics (as the bispectrum of the temperature
anisotropies) and to which the phenomenological study performed
in Ref. \cite{ks} applies.
A number of present and future CMB experiments, such as 
{\it WMAP} \cite{k}
 and {\it Planck}, have enough resolution to either constrain or detect 
non-Gaussianity of CMB anisotropy data parametrized by $f_{\rm NL}$
with high precision \cite{ks}.


\section{Non-Gaussianity in the various scenarios}
\label{non-gauss}
In each of the scenarios described in the previous section  
it is possible to calculate the level of non-linearity in the gravitational 
potential $\phi=\phi^{(1)}+\frac{1}{2} \phi^{(2)}$ (or $\psi$) in the 
longitudinal (Poisson) gauge \cite{poisson}. For instance the gravitational 
potential $\phi$ can be expressed in momentum space as
(up to a momentum-dependent function which has to be added to account for 
the condition $\langle \phi\rangle =0$) 
\begin{eqnarray}
\label{phimomspace}
\phi({\bf k}) &=& \phi^{(1)}({\bf k})+ 
\frac{1}{(2\pi)^3}
\int\, d^3 k_1\,d^3 k_2\, \delta^{(3)}\left({\bf k}_1+{\bf k}_2-{\bf k}\right)
\nonumber\\
&\times&
f^\phi_{\rm NL}\left({\bf k}_1,{\bf k}_2\right)
\phi^{(1)}({\bf k}_1)\phi^{(1)}({\bf k}_2)\, ,
\label{gauss}
\end{eqnarray}
where we have defined an effective ``momentum-dependent'' non-linearity 
parameter $f^\phi_{\rm NL}$. 
Here the linear lapse function
$\phi^{(1)}=\psi^{(1)}$ is a Gaussian random field. 
The gravitational potential bispectrum
reads

\begin{eqnarray}
&&\langle \phi({\bf k}_1) \phi({\bf k}_2) \phi({\bf k}_3)
\rangle=(2\pi)^3\,\delta^{(3)}\left({\bf k}_1+{\bf k}_2+{\bf k}_3\right)
\nonumber\\
&\times& \left[2\,f^\phi_{\rm NL}
\left({\bf k}_1,{\bf k}_2\right)\,
{\cal P}_\phi(k_1){\cal P}_\phi(k_2)+{\rm cyclic}\right]\, ,
\end{eqnarray}
where ${\cal P}_\phi(k)$ is the power-spectrum of the gravitational
potential. We stress again  that $f^{\phi}_{\rm NL}$ defines 
the non-Gaussianity of the gravitational potential, and it does not define 
the non-Gaussianity level of the CMB temperature fluctuations, see
Eq. (\ref{deltaT/T}).

We now give the predictions for the non-Gaussianity 
in the three scenario considered.

\subsection{Standard scenario}
In the one-single field model of inflation the initially tiny non-linearity 
in the cosmological  perturbations generated during the inflationary epoch
\cite{noi,maldacena} gets enhanced in the post-inflationary stages giving 
rise to a well-defined prediction for the non-linearity in the gravitational 
potentials. 
In Ref.~\cite{BMR} it has been shown how to calculate 
the second-order gravitational potential at second-order $\phi^{(2)}$ 
from an inflationary epoch to the 
radiation and matter dominated epochs by exploiting the conservation on large 
scales of the curvature perturbation $\zeta^{(2)}$ during
the inflationary stage and the radiation/matter phases.

As we have proved in Section V.A, the curvature perturbation $\zeta^{(2)}$
is conserved on large scales even during the reheating phase after 
inflation. Using the explicit expression for $\zeta^{(2)}$, the
second-order energy constraint (Eqs. (4.7) and (4.2) of Ref. \cite{BMR})
and the traceless part of the ($i$-$j$)-components of Einstein's equations 
at second order (Eq. (43) of Ref. \cite{BMR2}) we obtain, for perturbations 
re-entering the horizon during the matter-dominated era\footnote{Note that
this result coincides with the one given in Ref. \cite{BMR} up to an 
integrated over time gradient term which, however, is fully negligible when 
evaluating the bispectrum of the gravitational potential on large scales.},  
\be
\label{ee1}
f^\phi_{\rm NL}({\bf k}_1,{\bf k}_2) \simeq - \frac{1}{2} + 
g({\bf k}_1,{\bf k}_2)
\;,
\ee
where 
\be
g({\bf k}_1,{\bf k}_2) = \frac{{\bf k}_1 \cdot {\bf k}_2}
{k^2}\left(1+ 3\frac{{\bf k}_1\cdot {\bf k}_2}{k^2} \right)
\, ,
\ee
where ${\bf k} ={\bf k}_1 +  {\bf k}_2$. 
Notice that in the final bispectrum expression, the diverging terms arising 
from the infrared behaviour of $f^\phi_{\rm NL}({\bf k}_1,{\bf k}_2)$ 
are automatically regularized once the monopole term is subtracted from the 
definition of $\phi$ (by requiring that $\langle \phi \rangle$=0). 
Using Eq. (\ref{ppp}), we conclude that in the standard scenario where 
cosmological perturbations are generated by the inflaton field, 
the value of non-Gaussianity is provided by
\be
\label{ee1}
f_{\rm NL}({\bf k}_1,{\bf k}_2) \simeq - \frac{4}{3} + 
g({\bf k}_1,{\bf k}_2)
\,.
\ee

\subsection{Curvaton scenario}
In Ref.~\cite{BMR2} the 
level of non-Gaussianity in the gravitational potential 
$\phi$ has been calculated following in the sudden decay approximation 
the evolution of the gauge-invariant
curvature perturbation $\zeta^{(2)}$ produced by the initial isocurvature 
perturbations in the curvaton field $\sigma$. In the curvaton scenario we 
find 
\begin{equation}
\label{ee}
f^\phi_{\rm NL}=\left[\frac{7}{6}+\frac{5}{6}r-\frac{5}{4 r}\right] 
+  g({\bf k}_1,{\bf k}_2) \, .
\end{equation}
Using expression (\ref{ppp}), we find that the level of non-Gaussianity
in the curvaton scenario is provided by
\begin{equation}
f_{\rm NL}=\left[\frac{1}{3}+\frac{5}{6}r-\frac{5}{4 r}\right] 
+ g({\bf k}_1,{\bf k}_2) \, .
\end{equation}

\subsection{Inhomogeneous reheating scenario}
Using the technique developed in Refs.~\cite{BMR,BMR2} we can now calculate
the non linear parameter $f^{\phi}_{\rm NL}$ 
in the inhomogeneous reheating scenario.
We can switch from the spatially flat gauge to the Poisson gauge, 
since the results obtained in the previous section 
are gauge-invariant, involving the curvature perturbations. 
This is evident, for example, from Eq.~(\ref{risfin}).

During the matter dominated era from Eq.~(\ref{qqq}) it turns out that 
\cite{BMR}
\bea
\label{zetam}
\zeta^{(2)}&=&-\psi^{(2)}+\frac{1}{3}\frac{\delta^{(2)}\rho}{\rho}
+\frac{5}{9} \left( \frac{\delta^{(1)}\rho}{\rho}\right)^2 \nonumber\\
&=&-\psi^{(2)}+\frac{1}{3}\frac{\delta^{(2)}\rho}{\rho}+\frac{20}{9} 
\left(\psi^{(1)} \right)^2\, , 
\eea
where in the last step we have used that on large scales 
$\delta^{(1)}\rho/\rho=-2\psi^{(1)}$ in the Poisson gauge~\cite{BMR}.
Eq.~(\ref{zetam}) combined with Eq.~(\ref{risfin}) which gives the constant 
value on large scales of the curvature perturbation $\zeta^{(2)}$ during the 
matter dominated era, yields
\be
\psi^{(2)}-\frac{1}{3}\frac{\delta^{(2)}\rho}{\rho}=\frac{5}{6} 
\left( \psi^{(1)} \right)^2\, ,
\ee
where we have used the usual relation between 
the curvature perturbation and the Bardeen potential $\zeta^{(1)}=
-\frac{5}{3} \psi^{(1)}$ during the matter dominated era. 
From the (0-0) and ($i$-$j$)-components of Einstein equations, see Eqs.~(A.39) 
and (A.42-43) in Ref.~\cite{noi}, the following  
relations hold on large scales during the matter dominated epoch
\bea
\phi^{(2)}&=&-\frac{1}{2}\frac{\delta^{(2)}\rho}{\rho}+4\left( \psi^{(1)} 
\right)^2\, , \nonumber\\
\psi^{(2)}-\phi^{(2)}&=& - \frac{2}{3}\left( \psi^{(1)} \right)^2
+ \frac{10}{3}\nabla^{-2}\left(\psi^{(1)}\nabla^2\psi^{(1)}\right)\nonumber\\
&-&10\,\nabla^{-4}\left(\partial^i\partial_j\left(\psi^{(1)}\partial_i
\partial^j \psi^{(1)}\right)\right)\
\, .
\eea
We then read the 
non-linearity parameter for the 
gravitational potential $\phi=\phi^{(1)}+\frac{1}{2}\phi^{(2)}$
\be
f^\phi_{\rm NL}=\frac{3}{4} + g({\bf k}_1,{\bf k}_2) \, .
\ee

Using expression (\ref{ppp}), we find that the level of non-Gaussianity
in the inhomogeneous reheating  scenario where $\Gamma\propto
\chi^2$  is provided by

\be
f_{\rm NL}= - \frac{1}{12} + g({\bf k}_1,{\bf k}_2) \, .
\ee

\subsection{Some comments}
At this point, we would like to make some comments on the  
determination
of the non-linear parameter performed for the curvaton scenario in Ref.
\cite{LUW}
and for the inhomogeneous scenario in Refs. \cite{gamma1,gamma2}. Our results
differ from those previously obtained for several
reasons. First, the value of $f_{\rm NL}$ has been obtained in Refs.
\cite{LUW,gamma1,gamma2} by extending at second-order the first-order relation
$\phi^{(1)}=-\frac{3}{5}\zeta^{(1)}$. This procedure is not correct
since this linear relation is lost at second-order. Secondly, the value of
$\zeta^{(2)}$ has been obtained for the
curvaton scenario in Ref. \cite{LUW}
by merely expanding the energy-density fluctuation $\rho$ at second-order and
in the inhomogeneous scenario \cite{gamma1,gamma2} by expanding
the decay rate $\Gamma$ at second-order. However, the gauge-invariant
curvature perturbation contains additional second-order terms
of the form $({\rm first-order})^2$, see Eq. (\ref{qqq}).
Finally, the value of the non-linear parameter $f_{\rm NL}$ has to be
defined through the exact relation between the
temperature anisotropies and the (gauge-invariant) gravitational potentials,
see Eq. (\ref{deltaT/T}).


\section{Conclusions}
In this paper we have presented a general gauge-invariant
formalism to study the evolution of curvature perturbations at second-order.
In particular, we have addressed the evolution of the
total curvature perturbation in the three scenarios for the
generation of the cosmological perturbations on large scales: the
standard scenario where perturbations are induced by the inflaton
field and the curvaton and the inhomogeneous scenarios where 
the curvature perturbation is produced by initial isocurvature
perturbations. We have calculated the exact formula for the
second-order temperature fluctuations on large scales extending
the well-known expression for the Sachs-Wolfe effect at first-order.
We have provided the proper definition of the non-linear parameter
entering in the determination of the bispectrum of temperature
anisotropies, thus clarifying what is the quantity to be used in
the analysis aimed to look for non-Gaussian properties in the
CMB anisotropies such as the one given in Ref. \cite{ks}.
Finally, we have computed the values of non-Gaussianity in the
various scenarios for the generation of the cosmological
perturbations.
According to Ref. \cite{ks}, the minimum value of
$\vert f_{\rm NL}\vert$ that will become detectable from the analysis of
{\it WMAP} and {\it Planck} data, 
after properly subtracting detector noise and
foreground contamination, is about  $20$ and $5$, respectively, 
and 3 for an ideal experiment \cite{ks}.
Our findings imply that a future determination of non-Gaussianity 
at a level much larger than $f_{\rm NL}\sim 5$ would favour
the curvaton scenario, while the value of non-Gaussianity
provided by the standard scenario is below the
minimum  value $\sim 5$ advocated in Ref. \cite{ks}. However, such a lower 
bound has been obtained by analysing the capability of the bispectrum 
statistics to detect non-Gaussianity of the type discussed in this paper. 
Alternative statistical estimators based on the multivariate 
distribution function of the spherical harmonics of CMB maps \cite{rome}, 
or wavelet-based analyses \cite{wavelet}, together with a systematic 
application of MonteCarlo-simulated non-Gaussian CMB maps \cite{maps}, 
may allow to achieve sensitivity to lower values of $f_{\rm NL}$. 

\vskip 0.5truecm

\centerline{\bf Acknowledgments}

N.B. thanks PPARC for financial support. 



\begin{references}

\bibitem{guth81} A. Guth, Phys. Rev. D {\bf 23}, 347 (1981)

\bibitem{lrreview} D. H. Lyth and A. Riotto, Phys.~Rept.~314 1 (1999);
A.~Riotto, hep-ph/0210162; W.~H.~Kinney, astro-ph/0301448.

\bibitem{muk81} V. F. Mukhanov and G. V. Chibisov, JETP  Lett. {\bf 33}, 532 
(1981).

\bibitem{hawking82} S. W. Hawking, Phys. Lett. {\bf 115B}, 295 (1982).

\bibitem{starobinsky82} A. Starobinsky, Phys. Lett. {\bf 117B}, 175 (1982).

\bibitem{guth82} A. Guth and S. Y. Pi, Phys. Rev. Lett. {\bf 49}, 1110 (1982).

\bibitem{bardeen83} J. M. Bardeen, P. J. Steinhardt, and M. S. Turner, Phys. 
Rev. D {\bf 28}, 679 (1983).

\bibitem{smoot92} G. F. Smoot {\it et al.} Astrophys. J. {\bf 396}, L1 (1992).

\bibitem{bennett96} C. L. Bennett {\it et al.} Astrophys. J. {\bf 464}, L1 
(1996).

\bibitem{gorski96} K. M. Gorski {\it et al.} Astrophys. J. {\bf 464}, L11 
(1996).

\bibitem{wmap1} C.~L.~Bennett {\it et al.},
Astrophys.\ J.\ Suppl.\  {\bf 148}, 1 (2003)

\bibitem{ex} H.~V.~Peiris {\it et al.},
Astrophys.\ J.\ Suppl.\  {\bf 148}, 213 (2003); 
V.~Barger, H.~S.~Lee and D.~Marfatia, Phys.\ Lett.\ B {\bf 565}, 33 (2003)
[arXiv:hep-ph/0302150]; 
W.~H.~Kinney, E.~W.~Kolb, A.~Melchiorri and A.~Riotto,
arXiv:hep-ph/0305130; 
S.~M.~Leach and A.~R.~Liddle,
arXiv:astro-ph/0306305.

\bibitem{curvaton1} K.~Enqvist and M.~S.~Sloth,
Nucl.\ Phys.\ B {\bf 626}, 395 (2002)
[arXiv:hep-ph/0109214].

\bibitem{LW}
D.~H.~Lyth and D.~Wands,
Phys.\ Lett.\ B {\bf 524}, 5 (2002).

\bibitem{curvaton3} T.~Moroi and T.~Takahashi,
Phys.\ Lett.\ B {\bf 522}, 215 (2001)
[Erratum-ibid.\ B {\bf 539}, 303 (2002)]
[arXiv:hep-ph/0110096].

\bibitem{LUW}
D.~H.~Lyth, C. Ungarelli and D.~Wands,  
Phys.~Rev.D {\bf 67}, 023503 (2003).

\bibitem{gamma1} G.~Dvali, A.~Gruzinov and M.~Zaldarriaga,
arXiv:astro-ph/0303591.

\bibitem{gamma2} 
L.~Kofman,
arXiv:astro-ph/0303614.

\bibitem{gamma3}
G.~Dvali, A.~Gruzinov and M.~Zaldarriaga,
arXiv:astro-ph/0305548.

\bibitem{Qreh}                  
T.~Hamazaki and H.~Kodama, Prog.Theor.Phys.~{\bf 96}, 1123 (1996). 

\bibitem{Mollerach}
S.~Mollerach,
Phys.\ Rev.\ D {\bf 42} (1990) 313.

\bibitem{MFB}
V.~F.~Mukhanov, H.~A.~Feldman and R.~H.~Brandenberger,
Phys.\ Rept.\  {\bf 215} (1992) 203.

\bibitem{GBW}
J.~Garcia-Bellido and D.~Wands,
Phys.\ Rev.\ D {\bf 53}, 5437 (1996) [arXiv:astro-ph/9511029].

\bibitem{WMLL}
D.~Wands, K.~A.~Malik, D.~H.~Lyth and A.~R.~Liddle,
Phys.\ Rev.\ D {\bf 62} (2000) 043527 [astro-ph/0003278].

\bibitem{wmu} K.~A.~Malik, D.~Wands and C.~Ungarelli,
Phys.\ Rev.\ D {\bf 67}, 063516 (2003)
[arXiv:astro-ph/0211602].

\bibitem{k}
E.~Komatsu {\it et al.}, Astrophys.\ J.\ Suppl.\  {\bf 148}, 119 (2003)
[arXiv:astro-ph/0302223]. 

\bibitem{ks}
E.~Komatsu and D.~N.~Spergel,
Phys.\ Rev.\ D {\bf 63}, 063002 (2001).

\bibitem{Turner}
M.~S.~Turner,
Phys.\ Rev.\ D {\bf 28} (1983) 1243.

\bibitem{MatRio}
S.Matarrese and A.~Riotto, JCAP 0308, 007 (2003).

\bibitem{KSa}
H.~Kodama and M.~Sasaki,
Prog.\ Theor.\ Phys.\ Suppl.\  {\bf 78} (1984) 1.

\bibitem{MMB} S.~Matarrese, S.~Mollerach and M.~Bruni, Phys.\ Rev.\ {\bf D58},
043504 (1998).

\bibitem{noi} V.~Acquaviva, N.~Bartolo, S.~Matarrese and A.~Riotto,
ucl.\ Phys.\ B {\bf 667}, 119 (2003) [arXiv:astro-ph/0209156]. 

\bibitem{mw} K.~A.~Malik and D.~Wands,
arXiv:astro-ph/0307055.

\bibitem{BMR}
N.~Bartolo, S.~Matarrese and A.~Riotto, arXiv:astro-ph/0308088.

\bibitem{BMR2}
N.~Bartolo, S.~Matarrese and A.~Riotto, arXiv:hep-ph/0309033.

\bibitem{poisson} 
C.~P.~Ma and E.~Bertschinger,
Astrophys.\ J.\  {\bf 455}, 7 (1995).

\bibitem{molmat}
S.~Mollerach and S.~Matarrese,
Phys.\ Rev.\ D {\bf 56}, 4494 (1997).

\bibitem{pyne}
T.~Pyne and S.~M.~Carroll,
Phys.\ Rev.\ D {\bf 53}, 2920 (1996).

\bibitem{rs} X.~c.~Luo and D.~N.~Schramm,
Phys.\ Rev.\ Lett.\  {\bf 71}, 1124 (1993); 
D.~Munshi, T.~Souradeep and A.~A.~Starobinsky,
Astrophys.\ J.\  {\bf 454}, 552 (1995); 
S.~Mollerach, A.~Gangui, F.~Lucchin and S.~Matarrese,
Astrophys.\ J.\  {\bf 453}, 1 (1995); 
D.~N.~Spergel and D.~M.~Goldberg,
Phys.\ Rev.\ D {\bf 59}, 103001 (1999); 
A.~Cooray,
Phys.\ Rev.\ D {\bf 65}, 083518 (2002).

\bibitem{kphd} E.~Komatsu,
arXiv:astro-ph/0206039.

\bibitem{maldacena}
J.~Maldacena,
JHEP {\bf 0305}, 013 (2003).

\bibitem{rome}
F.~K.~Hansen, D.~Marinucci and N.~Vittorio,
Phys.\ Rev.\ D {\bf 67}, 123004 (2003).

\bibitem{wavelet} See, {\it e.g.}, 
N.~Aghanim, M.~Kunz, P.~G.~Castro and O.~Forni,
Astron.\ Astrophys.\  {\bf 406} (2003) 797;
L.~Cayon, E.~Martinez-Gonzalez, F.~Argueso, A.~J.~Banday and K.~M.~Gorski,
Mon.\ Not.\ Roy.\ Astron.\ Soc.\  {\bf 339}, 1189 (2003)

\bibitem{maps} 
E.~Komatsu, D.~N.~Spergel and B.~D.~Wandelt,
arXiv:astro-ph/0305189; M.~Liguori, S.~Matarrese and L.~Moscardini,
arXiv:astro-ph/0306248, to be published in Astrophys.\ J. 
















\end{references}
\end{document}